\documentclass[prb,preprint,onecolumn,showpacs,preprintnumbers,amsmath,amssymb,superscriptaddress]{revtex4-2}

\usepackage{graphicx}
\usepackage{dcolumn}
\usepackage{bm}
\usepackage{amsthm}
\usepackage{amsmath}
\usepackage{amssymb}
\usepackage{xcolor}

\begin{document}

\title{Micromagnetic simulation of neutron scattering from spherical nanoparticles:~Effect of pore-type defects}

\author{Evelyn Pratami Sinaga}\email[Electronic address: ]{evelyn.sinaga@uni.lu}
\author{Michael P.\ Adams}
\author{Mathias Bersweiler}
\author{Laura G.\ Vivas}
\author{Eddwi H.\ Hasdeo}
\affiliation{Department of Physics and Materials Science, University of Luxembourg, 162A~Avenue de la Faiencerie, L-1511 Luxembourg, Grand Duchy of Luxembourg}
\author{Jonathan Leliaert}
\affiliation{Department of Solid State Sciences, Ghent University, Krijgslaan 281/S1, 9000 Ghent, Belgium}
\author{Philipp Bender}
\affiliation{Heinz Maier-Leibnitz Zentrum, Technische Universität M\"unchen, D-85748 Garching, Germany}
\author{Dirk Honecker}
\affiliation{ISIS Neutron and Muon Source, Rutherford Appleton Laboratory, Didcot, United Kingdom}
\author{Andreas Michels}\email[Electronic address: ]{andreas.michels@uni.lu}
\affiliation{Department of Physics and Materials Science, University of Luxembourg, 162A~Avenue de la Faiencerie, L-1511 Luxembourg, Grand Duchy of Luxembourg}

\date{\today}

\begin{abstract}
We employ micromagnetic simulations to model the effect of pore-type microstructural defects on the magnetic small-angle neutron scattering cross section and the related pair-distance distribution function of spherical magnetic nanoparticles. Our expression for the magnetic energy takes into account the isotropic exchange interaction, the magnetocrystalline anisotropy, the dipolar interaction, and an externally applied magnetic field. The signatures of the defects and the role of the dipolar energy are highlighted and the effect of a particle-size distribution is studied. The results serve as a guideline to the experimentalist.
\end{abstract}

\maketitle

\section{Introduction}

Magnetic small-angle neutron scattering (SANS) is the method of choice for studying spin structures on a mesoscopic length scale of typically $1$$-$$100 \, \mathrm{nm}$ and inside the volume of magnetic materials~\cite{rmp2019,michelsbook}. A growing number of experimental investigations on \textit{magnetic nanoparticles}, in particular using polarized neutrons, unanimously suggest that the encountered spin configurations are highly complex and exhibit a variety of nonuniform, canted, or core-shell-type textures (see, e.g., Refs.~\cite{michels08epl,disch2012,kryckaprl2014,ijiri2014,guenther2014,maurer2014,dennis2015,grutter2017,oberdick2018,krycka2019,benderapl2019,bersweiler2019,zakutna2020,dirkreview2022} and references therein). However, a problem arises since the prototypical magnetic SANS data analysis is largely based on structural form-factor-type models for the cross section. These are borrowed from nuclear SANS and do not properly account for the existing spin inhomogeneity inside magnetic nanoparticles. On the other hand, analytical as well numerical computations of the magnetic SANS cross section~\cite{michels2013,michels2014jmmm,mettus2015,erokhin2015,metmi2015,metmi2016,michelsPRB2016,michelsdmi2019,mistonov2019,malyeyev2022,metlov2022} strongly suggest that for the analysis of experimental magnetic SANS data the spatial nanometer scale variation of the orientation and magnitude of the magnetization vector field must be taken into account; and that macrospin-based models---assuming a \textit{uniform} magnetization---are not adequate.

Theoretical descriptions of magnetic SANS are based on Brown's static equations of micromagnetics~\cite{brown}, which are a set of nonlinear partial differential equations for the magnetization along with complex boundary conditions on the sample's surface~\cite{evelynprbcomment2022}. Therefore, closed-form analytical results for the SANS cross section are restricted to special limiting cases such as the approach-to-saturation regime, where the governing equations can be linearized~\cite{michels2013,metmi2015,metmi2016,michelsPRB2016,malyeyev2022,metlov2022}. Recently, we have carried out numerical micromagnetic computations to study the magnetic SANS cross section of microstructural-defect-free spherical nanoparticles during their transition from the single-domain to the multi-domain state~\cite{laura2017,laura2020}. The results for the magnetic SANS signal and correlation function have revealed pronounced differences as compared to the superspin model and provided guidance for the experimentalist to identify nonuniform vortex-type spin structures inside nanoparticles.

In this work, we extend the numerical micromagnetic approach to include the effects of microstructural pore-type defects and of a particle-size distribution function. Defects in nanoparticles (e.g., surface anisotropy, vacancies, antiphase boundaries) are known for a long time to give rise to spin disorder and in this way influence the macroscopic magnetic properties (see, e.g., Refs.~\cite{garanin2003,nedelkoski2017,lappas2019,zakutna2020,lakbenderdisch2021}). Therefore, finding their signature in the magnetic SANS cross section and correlation function is highly desirable.

The article is organized as follows:~In Sec.~\ref{msanssec}, we display the expressions for the magnetic SANS cross section and for the pair-distance distribution function. In Sec.~\ref{mumagdetails}, we provide information on the micromagnetic simulations and on the implementation of the microstructural defects. In Sec.~\ref{ressec}, we present and discuss the results, with Sec.~\ref{resseca} focusing on the real-space spin structures, the magnetization, and the SANS observables and Sec.~\ref{ressecb} discussing the effect of a particle-size distribution function. Finally, Sec.~\ref{summary} summarizes the main findings of this study and provides an outlook on future challenges.

\section{Magnetic SANS cross section and pair-distance distribution function}
\label{msanssec}

\begin{figure}[tb!]
\centering
\resizebox{0.75\columnwidth}{!}{\includegraphics{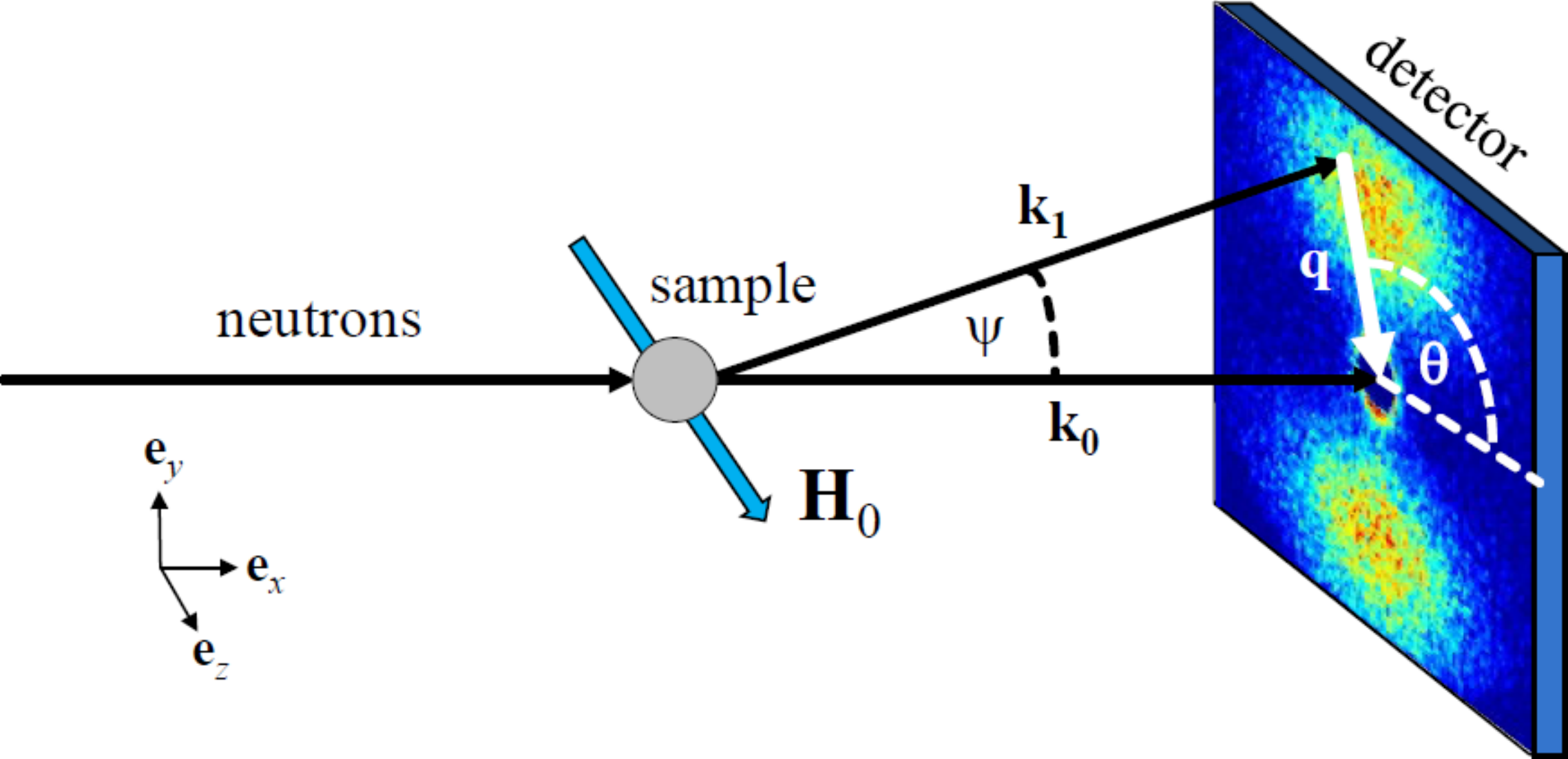}}
\caption{Sketch of the scattering geometry assumed in the micromagnetic simulations. The applied magnetic field $\mathbf{H}_0 \parallel \mathbf{e}_z$ is perpendicular to the wave vector $\mathbf{k}_0 \parallel \mathbf{e}_x$ of the incident neutron beam ($\mathbf{k}_0 \perp \mathbf{H}_0$). The momentum-transfer or scattering vector $\mathbf{q}$ is defined as the difference between $\mathbf{k}_0$ and $\mathbf{k}_1$, i.e., $\mathbf{q} = \mathbf{k}_0 - \mathbf{k}_1$. SANS is usually implemented as elastic scattering ($k_0 = k_1 = 2\pi / \lambda$), and the component of $\mathbf{q}$ along the incident neutron beam, here $q_x$, is much smaller than the other two components, so that $\mathbf{q} \cong \{ 0, q_y, q_z \} = q \{ 0, \sin\theta, \cos\theta \}$. This demonstrates that SANS probes predominantly correlations in the plane perpendicular to the incident beam. The angle $\theta = \angle(\mathbf{q}, \mathbf{H}_0)$ is used to describe the angular anisotropy of the recorded scattering pattern on the two-dimensional position-sensitive detector. For elastic scattering, the magnitude of $\mathbf{q}$ is given by $q = (4\pi / \lambda) \sin(\psi/2)$, where $\lambda$ denotes the mean wavelength of the neutrons and $\psi$ is the scattering angle.}
\label{fig1}
\end{figure}

The quantity of interest is the elastic magnetic differential scattering cross section $d \Sigma_{\mathrm{M}} / d \Omega$, which is usually recorded on a two-dimensional position-sensitive detector. For the most commonly used scattering geometry in magnetic SANS experiments, where the applied magnetic field $\mathbf{H}_0 \parallel \mathbf{e}_z$ is perpendicular to the wave vector $\mathbf{k}_0 \parallel \mathbf{e}_x$ of the incident neutrons (see Fig.~\ref{fig1}), $d \Sigma_{\mathrm{M}} / d \Omega$ (for unpolarized neutrons) can be written as~\cite{rmp2019}:
\begin{eqnarray}
\frac{d \Sigma_{\mathrm{M}}}{d \Omega} = \frac{8 \pi^3}{V} b_{\mathrm{H}}^2 \left( |\widetilde{M}_x|^2 + |\widetilde{M}_y|^2 \cos^2\theta \right. \nonumber \\ \left. + |\widetilde{M}_z|^2 \sin^2\theta - (\widetilde{M}_y \widetilde{M}_z^{\ast} + \widetilde{M}_y^{\ast} \widetilde{M}_z) \sin\theta \cos\theta \right) ,
 \label{eq:Eq.1}
\end{eqnarray}
where $V$ is the scattering volume, $b_{\mathrm{H}} = 2.91 \times 10^8 \, \mathrm{A}^{-1}\mathrm{m}^{-1}$ is the magnetic scattering length in the small-angle regime (the atomic magnetic form factor is approximated by $1$, since we are dealing with forward scattering), $\widetilde{\mathbf{M}}(\mathbf{q}) = \{ \widetilde{M}_x(\mathbf{q}), \widetilde{M}_y(\mathbf{q}), \widetilde{M}_z(\mathbf{q}) \}$ represents the Fourier transform of the magnetization vector field $\mathbf{M}(\mathbf{r}) = \{ M_x(\mathbf{r}), M_y(\mathbf{r}), M_z(\mathbf{r}) \}$, $\theta$ denotes the angle between $\mathbf{q}$ and $\mathbf{H}_0$, and the asterisk ``$*$'' marks the complex-conjugated quantity. Note that in the perpendicular scattering geometry the Fourier components are evaluated in the plane $q_x = 0$ (compare Fig.~\ref{fig1}). For a uniformly magnetized spherical particle with its saturation direction parallel to $\mathbf{e}_z$, i.e., $M_x = M_y = 0$, Eq.~(\ref{eq:Eq.1}) reduces to:
\begin{eqnarray}
\frac{d \Sigma_{\mathrm{M}}}{d \Omega} = V_{\mathrm{p}} (\Delta \rho)_{\mathrm{mag}}^2 \, 9 \left( \frac{j_1(qR)}{qR} \right)^2 \sin^2\theta ,
\label{homomagsans2}
\end{eqnarray}
where $V_{\mathrm{p}} = \frac{4\pi}{3} R^3$ is the sphere volume, $(\Delta \rho)_{\mathrm{mag}}^2 = b_{\mathrm{H}}^2 \left( \Delta M \right)^2$ is the magnetic scattering-length density contrast, and $j_1(qR)$ is the first-order spherical Bessel function. The well-known analytical result for the homogeneous sphere case, Eq.~(\ref{homomagsans2}), and its correlation function [see Eq.~(\ref{pvonreq}) below] serve as a reference for comparison to the nonuniform case.

The pair-distance distribution function $p(r)$ can be computed from the azimuthally-averaged magnetic SANS cross section according to:
\begin{eqnarray}
\label{pvonreqintegral}
p(r) = r^2 \int\limits_0^{\infty} \frac{d \Sigma_{\mathrm{M}}}{d \Omega}(q) j_0(qr) q^2 dq ,
\end{eqnarray}
where $j_0(qr) = \sin(qr)/(qr)$ is the spherical Bessel function of zero order; $p(r)$ corresponds to the distribution of real-space distances between volume elements inside the particle weighted by the excess scattering-length density distribution; see the reviews by Glatter~\cite{glatterchapter} and by Svergun and Koch~\cite{svergun03} for detailed discussions of the properties of $p(r)$. Apart from constant prefactors, the $p(r)$ of the azimuthally-averaged single-particle cross section [Eq.~(\ref{homomagsans2})], corresponding to a uniform sphere magnetization, equals (for $r \leq 2R$):
\begin{eqnarray}
\label{pvonreq}
p(r) = r^2 \left( 1 - \frac{3r}{4R} + \frac{r^3}{16 R^3} \right) .
\end{eqnarray}
We also display the correlation function $c(r)$, which is related to $p(r)$ via
\begin{eqnarray}
\label{cvonreq}
c(r) = p(r)/r^2 .
\end{eqnarray}
As we will demonstrate in the following, when the particles' spin structure is inhomogeneous, the $d \Sigma_{\mathrm{M}} / d \Omega$ and the corresponding $p(r)$ and $c(r)$ differ significantly from the homogeneous case [Eqs.~(\ref{homomagsans2}) and (\ref{pvonreq})]. Due to the $r^2$~factor, features in $p(r)$ at medium and large distances are more pronounced than in $c(r)$.

The magnetic SANS cross section $d \Sigma_{\mathrm{M}} / d \Omega$ [Eq.~(\ref{eq:Eq.1})] depends on both the magnitude $q$ and the orientation $\theta$ of the scattering vector $\mathbf{q}$ on the two-dimensional detector. The origin of its angular anisotropy ($\theta$~dependence) is twofold~\cite{erokhin2015}: (i)~the trigonometric functions in Eq.~(\ref{eq:Eq.1}) are due to the dipolar interaction between the magnetic moment of the neutron and the magnetization of the sample, while (ii)~the Fourier components $\widetilde{M}_{x,y,z}$ may additionally depend on the angle $\theta$ via the intrinsic dipolar interaction between the magnetic moments comprising the sample. Variation of the external magnetic field strength changes the relative contributions of the $\widetilde{M}_{x,y,z}$ to $d \Sigma_{\mathrm{M}} / d \Omega$ and also their angular anisotropy [compare, e.g., Eq.~(\ref{homomagsans2})].

\section{Details on the micromagnetic simulations}
\label{mumagdetails}

The micromagnetic computations of the spin structure of a single spherical nanoparticle were performed using the GPU-based open-source software package MuMax3 (version~3.10)~\cite{mumax3new,leliaert2019}. This code allows the calculation of the space and time-dependent magnetization of nano- and micron-sized ferromagnets. MuMax3 is based on a finite-difference discretization scheme of space using a two-dimensional or three-dimensional grid of orthorhombic cells. In the micromagnetic simulations we have taken into account all four standard contributions to the total magnetic Gibbs free energy, i.e., Zeeman energy $E_{\mathrm{Z}}$ in the external magnetic field $\mathbf{H}_0$, dipolar (magnetostatic) interaction energy $E_{\mathrm{D}}$, energy of the (cubic) magnetocrystalline anisotropy $E_{\mathrm{ani}}$, and isotropic and symmetric exchange energy $E_{\mathrm{ex}}$. The expressions for these energies are the following:
\begin{eqnarray}
E_{\mathrm{Z}} = - \mu_0 \int_V \mathbf{M} \cdot \mathbf{H}_0 \, dV ,
\end{eqnarray}
\begin{eqnarray}
E_{\mathrm{D}} = - \frac{1}{2} \mu_0 \int_V \mathbf{M} \cdot \mathbf{H}_{\mathrm{D}} \, dV ,
\end{eqnarray}
\begin{eqnarray}
E_{\mathrm{ani}} = \int_V K_1 \left( m^2_x m^2_y + m^2_y m^2_z + m^2_x m^2_z \right) \, dV ,
\end{eqnarray}
\begin{eqnarray}
E_{\mathrm{ex}} = \int A \left[ (\nabla m_x)^2 + (\nabla m_y)^2 + (\nabla m_z)^2 \right] \, dV ,
\end{eqnarray}
where $\mathbf{m}(\mathbf{r}) = \mathbf{M}(\mathbf{r})/M_{\mathrm{s}}$ denotes the unit magnetization vector with $M_{\mathrm{s}}$ the saturation magnetization, $\mathbf{H}_0 \parallel \mathbf{e}_z$ is the (constant) applied magnetic field, $\mathbf{H}_{\mathrm{D}}(\mathbf{r})$ is the magnetostatic self-interaction field, $K_1$ is the first-order cubic anisotropy constant, $A$ is the exchange-stiffness constant, $\mu_0 = 4\pi \times 10^{-7} \, \mathrm{Tm/A}$, and the integrals are taken over the volume of the sample. In the simulations, we used the following material parameters for iron (Fe):~saturation magnetization $M_{\mathrm{s}} = 1700 \, \mathrm{kA/m}$, exchange-stiffness constant $A = 1.0 \times 10^{-11} \, \mathrm{J/m}$, and a first-order cubic anisotropy constant of $K_1 = 47 \, \mathrm{kJ/m^3}$. These values result in a critical single-domain diameter of $D_{\mathrm{cr}}^{\mathrm{sd}} \sim 72 \sqrt{A K_1}/(\mu_0 M_{\mathrm{s}}^2) = 13.6$~nm~\cite{kronfahn03}.

\begin{figure}[tb!]
\centering
\resizebox{0.40\columnwidth}{!}{\includegraphics{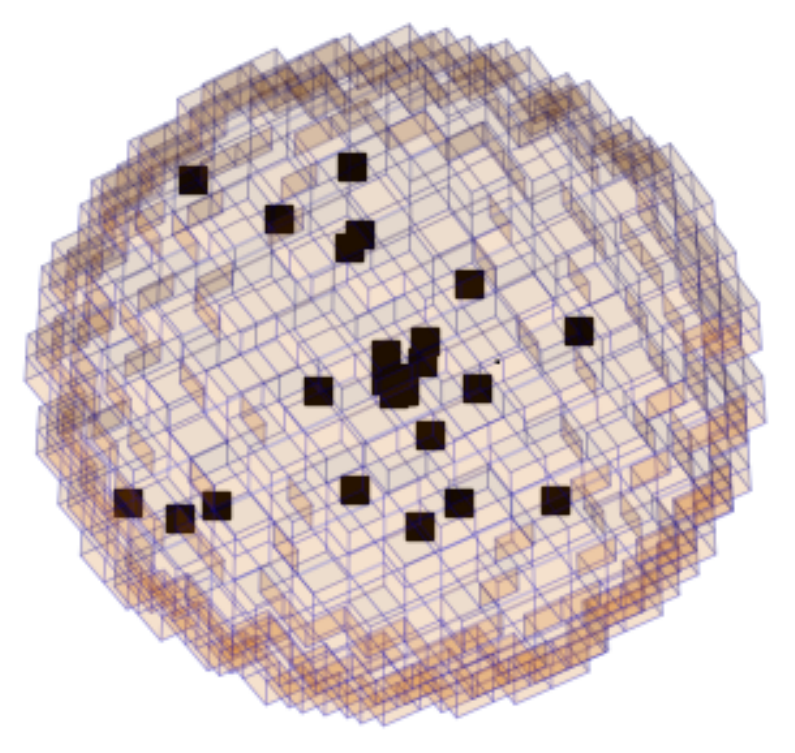}}
\caption{Discretization of the spherical simulation volume $V$ into cubical cells with a size of typically $2 \times 2 \times 2 \, \mathrm{nm^3}$ (particle size:~$D = 40 \, \mathrm{nm}$). The black squares mark the randomly-chosen ``defect'' cells with $M_{\mathrm{s}} = 0$.}
\label{fig2}
\end{figure}

Figure~\ref{fig2} displays the model used in the micromagnetic SANS simulations. The sphere volume is first discretized into cubical cells with a size of $2 \times 2 \times 2 \ \, \mathrm{nm}^3$ (finite-difference method). Next, we randomly assign defects into this structure, which are per definition cells with a saturation magnetization of $M_{\mathrm{s}} = 0$. This might model e.g.\ nonmagnetic pore-type defects. Assuming the size of a single atom to be $0.1 \, \mathrm{nm}$, such a hole comprises about 8000~atoms and represents, thus, a very strong perturbation in the magnetic nanoparticle structure. All simulations were carried out by first saturating the nanoparticle by a strong external field $\mathbf{H}_0$, and then the field was decreased (in steps of typically $5 \, \mathrm{mT}$) following the major hysteresis loop. For a given volume concentration of defects ($x_{\mathrm{d}} \sim 0$$-$$20 \, \%$), we perform, at each value of $H_0$, micromagnetic simulations for typically $N \sim 100$ random orientations between the magnetic easy axis of the particle and $\mathbf{H}_0$; for each random particle orientation, the defect distribution was randomly selected. All data shown in this paper, except the spin structures in Fig.~\ref{fig3}, correspond to an ensemble of randomly-oriented particles. Simulations on $10$-nm-sized single-domain particles (data not shown) yield the well-known values for the reduced remanence and coercivity predicted by the Stoner-Wohlfarth model~\cite{usov1997}.

For each step of $H_0$ and for each particular random easy-axis angle, we have obtained the equilibrium spin structure $M_{x,y,z}(x,y,z)$ by employing both the ``Relax'' and ``Minimize'' functions of MuMax3. The former solves the Landau-Lifshitz-Gilbert equation without the precessional term and the latter uses the conjugate-gradient method to find the configuration of minimum energy. The translational invariance of the grid obtained with the finite-difference method enables the usage of the fast Fourier transformation technique for the computation of the Cartesian Fourier components $\widetilde{M}_{x,y,z}(q_x,q_y,q_z)$ of $M_{x,y,z}(x,y,z)$. We have used the FFTW library~\cite{frigo2007} to compute and analyze the Fourier components of our nanoscopic magnetic configurations. These were then evaluated in the plane $q_x=0$ (corresponding to the scattering geometry shown in Fig.~\ref{fig1}) and used in Eq.~(\ref{eq:Eq.1}) to compute the magnetic SANS cross section $d \Sigma_{\mathrm{M}} / d \Omega$ according to:
\begin{eqnarray}
\frac{d \Sigma_{\mathrm{M}}}{d \Omega} = \sum_{i=1}^{N} \frac{d \Sigma_{\mathrm{M},i}}{d \Omega},
\label{sigmaaverage}
\end{eqnarray}
where $d \Sigma_{\mathrm{M},i} / d \Omega$ represents (for fixed $x_{\mathrm{d}}$ and $H_0$) the magnetic SANS cross section of a spherical particle with diameter $D$ and with a particular random easy-axis orientation ``$i$''. Equation~(\ref{sigmaaverage}) implies that interparticle-interference effects are ignored in the simulations. Likewise, the effect of temperature has also not been taken into account.

\section{Results and discussion \label{ressec}}

\subsection{Spin structure, magnetization, and magnetic SANS observables \label{resseca}}

\begin{figure*}[tb!]
\centering
\resizebox{1.0\columnwidth}{!}{\includegraphics{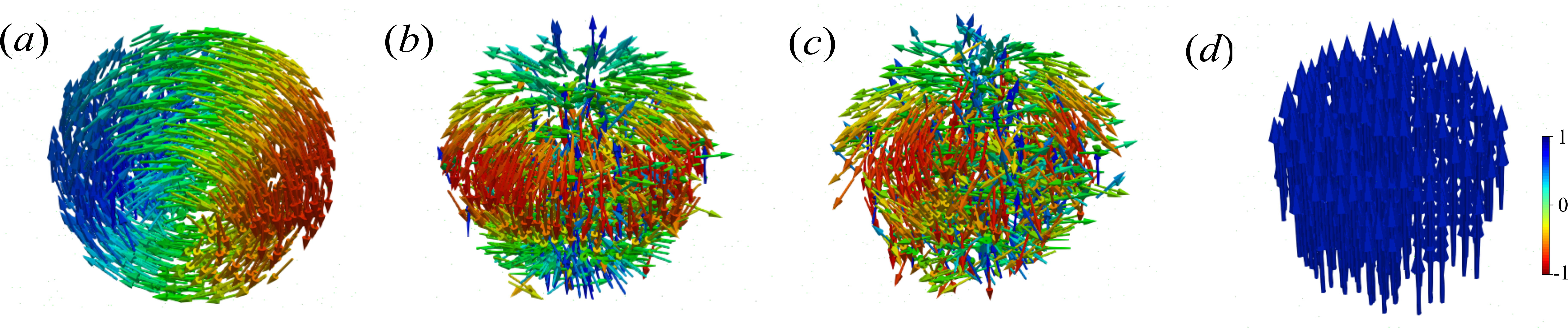}}
\caption{Spin structures (snap shots) of $40$-nm-sized Fe spheres with defect concentrations of (a)~$x_{\mathrm{d}} = 0 \, \%$, (b)~$x_{\mathrm{d}} = 5 \, \%$, and (c)~$x_{\mathrm{d}} = 15 \, \%$. (d)~Spin structure of a defect-free Fe sphere ($x_{\mathrm{d}} = 0 \, \%$) without taking into account the dipolar interaction in the energy minimization procedure ($E_{\mathrm{D}} = 0$). Applied magnetic field is $\mu_0 H_0 = 0.02 \, \mathrm{T}$ in (a)$-$(d); initially all structures were saturated. In the actual SANS simulations, the magnetic easy axis of the particle was randomly selected relative to $\mathbf{H}_0$. Here, in order to illustrate the effect of the defects, the easy axis was chosen the same in (a)$-$(d).}
\label{fig3}
\end{figure*}

Figure~\ref{fig3} depicts the spin structures of $40$-nm-sized Fe spheres for several defect concentrations $x_{\mathrm{d}}$ and at an applied magnetic field of $\mu_0 H_0 = 0.02$~T ($\mathbf{H}_0 \parallel \mathbf{e}_z$). While the defect-free case [Fig.~\ref{fig3}(a)] exhibits a vortex-type spin configuration (reproducing the results from~\cite{laura2020}), increasing $x_{\mathrm{d}}$ results in the progressive disordering (randomization) of the structure [Figs.~\ref{fig3}(b) and (c)]. For the here-considered particle size of $D = 40 \, \mathrm{nm}$, which is larger than the single-domain size of Fe ($D_{\mathrm{cr}}^{\mathrm{sd}} \cong 13.6 \, \mathrm{nm}$), the vortex structure in Fig.~\ref{fig3}(a) is clearly a consequence of the dipolar interaction. Leaving out the dipolar energy in the simulations (i.e., setting $E_{\mathrm{D}} = 0$) results in a quasi single-domain state [Fig.~\ref{fig3}(d)]. For $x_{\mathrm{d}} = 0 \, \%$ and small applied fields, the vortex structure appears for particle sizes $D \gtrsim 20 \, \mathrm{nm}$. Overall, we see that (for $D = 40 \, \mathrm{nm}$) adding defects changes the vortex-type spin structure significantly towards more disordered spin configurations.

\begin{figure}[tb!]
\centering
\resizebox{0.80\columnwidth}{!}{\includegraphics{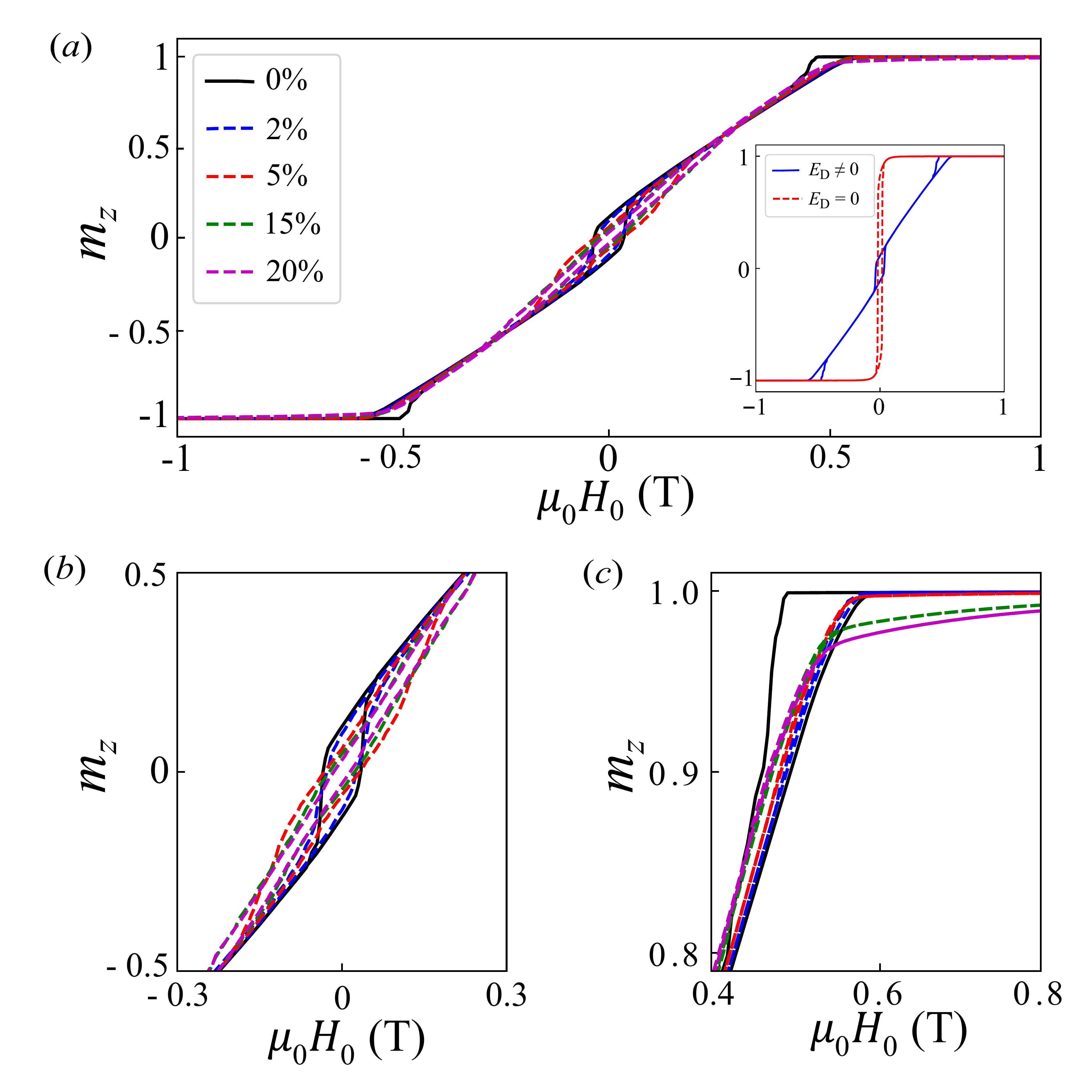}}
\caption{(a)~Normalized magnetization curves $m_z(H_0)$ of $40$-nm-sized Fe spheres for different defect concentrations $x_{\mathrm{d}}$ (see inset). (b)~Zoom of $m_z(H_0)$ around (a) $H_0 = 0$ and (c) around $\mu_0 H_0 = 0.6 \, \mathrm{T}$. The inset in (a) compares the magnetization curves of a defect-free $40$-nm sphere with and without the dipolar interaction energy $E_{\mathrm{D}}$.}
\label{fig4}
\end{figure}

In Fig.~\ref{fig4}, we show the reduced hysteresis curves $m_z(H_0) = M_z(H_0)/M_{\mathrm{s}}^{*}$ for various $x_{\mathrm{d}} \sim 0$$-$$20 \, \%$. We note that the magnetization is here normalized by the defect concentration, according to $M_{\mathrm{s}}^{*} = (1-x_{\mathrm{d}}) M_{\mathrm{s}}$. As expected, increasing $x_{\mathrm{d}}$ results in a reduction of the magnetization, in particular around the remanent state and in the approach-to-saturation regime [compare Figs.~\ref{fig4}(b) and (c)]. To be more quantitative, the reduced remanence decreases from $\sim 0.11$ at $x_{\mathrm{d}} = 0 \, \%$ to $\sim 0.04$ at $x_{\mathrm{d}} = 20 \, \%$, and the coercivity decreases from $35 \, \mathrm{mT}$ at $x_{\mathrm{d}} = 0 \, \%$ to $18 \, \mathrm{mT}$ at $x_{\mathrm{d}} = 20 \, \%$. The saturation field also increases from $\sim 0.48 \, \mathrm{T}$ for $x_{\mathrm{d}} = 0 \, \%$ to $\sim 0.53 \, \mathrm{T}$ for $x_{\mathrm{d}} = 20 \, \%$. The inset in Fig.~\ref{fig4}(a) depicts the effect of the dipolar interaction (for $x_{\mathrm{d}} = 0 \, \%$). For $E_{\mathrm{D}} = 0$, the shape of the $m_z(H_0)$ loop is rectangular and agrees with the predictions of the Stoner-Wohlfarth coherent-rotation model, i.e., we find a reduced remanence of $\sim 0.83$ and a coercivity of $0.33 \times 2 K_1/M_{\mathrm{s}} \cong 18 \, \mathrm{mT}$~\cite{usov1997}. With dipolar interaction, two regions with a large hysteresis are seen, one at around $0.5 \, \mathrm{T}$ [Fig.~\ref{fig4}(c)], which corresponds to the first deviation from the single-domain state, and one around the remanent state [Fig.~\ref{fig4}(a)], which corresponds to the nucleation of the vortex-type structure. In the following discussion of the SANS observables [$d\Sigma_{\mathrm{M}} / d\Omega$, $p(r)$, $c(r)$], we concentrate on the remanent state and on the high-field ``pocket'' at $0.5 \, \mathrm{T}$.

\begin{figure*}[tb!]
\centering
\resizebox{1.0\columnwidth}{!}{\includegraphics{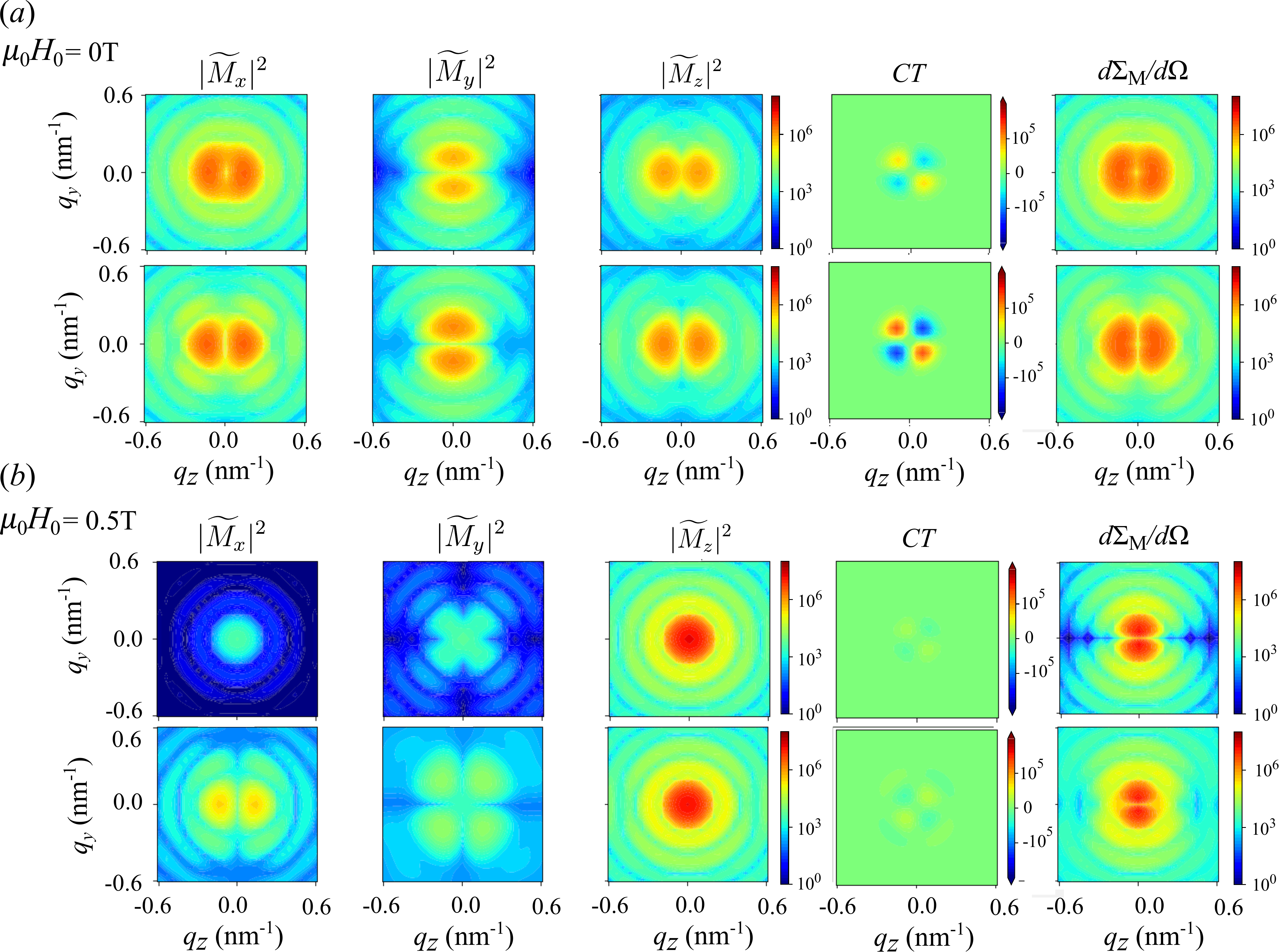}}
\caption{(a)~Two-dimensional Fourier components $|\widetilde{M}_x|^2$, $|\widetilde{M}_y|^2$, $|\widetilde{M}_z|^2$, $CT = - (\widetilde{M}_y \widetilde{M}_z^{\ast} + \widetilde{M}_y^{\ast} \widetilde{M}_z)$, and magnetic SANS cross section $d\Sigma_{\mathrm{M}} / d\Omega$ (in units of $\mathrm{cm}^{-1}$) of $40$-nm-sized Fe spheres at $\mu_0 H_0 = 0 \, \mathrm{T}$. (upper row) Defect concentration $x_{\mathrm{d}} = 0 \, \%$. (lower row) Defect concentration $x_{\mathrm{d}} = 15 \, \%$. $\mathbf{H}_0 \parallel \mathbf{e}_z$ is horizontal in the plane. Logarithmic color scales are used except for the $CT$s, which are displayed using a linear color scale. Note that the respective Fourier components are multiplied with the constant $8 \pi^3 V^{-1} b_H^2$ (in order to have the same unit as $d \Sigma_{\mathrm{M}} / d \Omega$), but not with the trigonometric functions in the expression for $d \Sigma_{\mathrm{M}} / d \Omega$ [compare Eq.~(\ref{eq:Eq.1})]. (b) Same as (a), but for $\mu_0 H_0 = 0.5 \, \mathrm{T}$.}
\label{fig5}
\end{figure*}

The two-dimensional magnetic SANS cross sections along with the Fourier components for $x_{\mathrm{d}} = 0 \, \%$ and $x_{\mathrm{d}} = 15 \, \%$ are shown in Fig.~\ref{fig5}(a) at the remanent state and in Fig.~\ref{fig5}(b) for $\mu_0 H_0 = 0.5 \, \mathrm{T}$. The $d\Sigma_{\mathrm{M}} / d\Omega$ in the remanent state are both horizontally elongated [Fig.~\ref{fig5}(a)], and the Fourier components exhibit only little variation with the defect concentration, e.g., regarding their angular anisotropy. The functions $|\widetilde{M}_x|^2$, $|\widetilde{M}_y|^2$, and $|\widetilde{M}_z|^2$ are (at $0 \, \mathrm{T}$) of comparable magnitude. Near saturation at $0.5 \, \mathrm{T}$ [Fig.~\ref{fig5}(b)], $d\Sigma_{\mathrm{M}} / d\Omega$ is dominated by the isotropic $|\widetilde{M}_z|^2$ Fourier component and exhibits the $\sin^2\theta$~anisotropy which is characteristic for an essentially saturated microstructure [compare Eq.~(\ref{eq:Eq.1})]. The contribution of the transversal Fourier components $|\widetilde{M}_x|^2$ and $|\widetilde{M}_y|^2$ to $d\Sigma_{\mathrm{M}} / d\Omega$ is much weaker than the longitudinal $|\widetilde{M}_z|^2$~contribution. When increasing the defect concentration $x_{\mathrm{d}}$ from $0 \, \%$ to $15 \, \%$, $|\widetilde{M}_x|^2$ at $0.5 \, \mathrm{T}$ changes its angular anisotropy, from isotropic to horizontally elongated, while the clover-leaf-type pattern of $|\widetilde{M}_y|^2$ becomes more pronounced. The $CT$s change their sign at the borders between the quadrants on the detector, e.g., in Fig.~\ref{fig5}(a) we see that $CT < 0$ for $0^{\circ} < \theta < 90^{\circ}$, $CT > 0$ for $90^{\circ} < \theta < 180^{\circ}$, and so on. We note that the $CT$ needs to be multiplied with $\sin\theta \cos\theta$ in  order to obtain the corresponding contribution to the magnetic SANS cross section [compare Eq.~(\ref{eq:Eq.1})]. We also emphasize that the $CT \sin\theta \cos\theta$~contribution to $d\Sigma_{\mathrm{M}} / d\Omega$ can be negative, in contrast to the other three contributions, which are strictly positive. Using the inequality $|\widetilde{M}_y \cos\theta - \widetilde{M}_z \sin\theta|^2 \geq 0$, it is easily seen that the contribution $CT \sin\theta \cos\theta$ is, however, always smaller than the sum of the other terms (as it must be). The results in Fig.~\ref{fig5} underline that, generally, the Fourier components in the magnetic SANS cross section [Eq.~(\ref{eq:Eq.1})] are anisotropic functions of the angle $\theta$.

\begin{figure*}[tb!]
\centering
\resizebox{0.75\columnwidth}{!}{\includegraphics{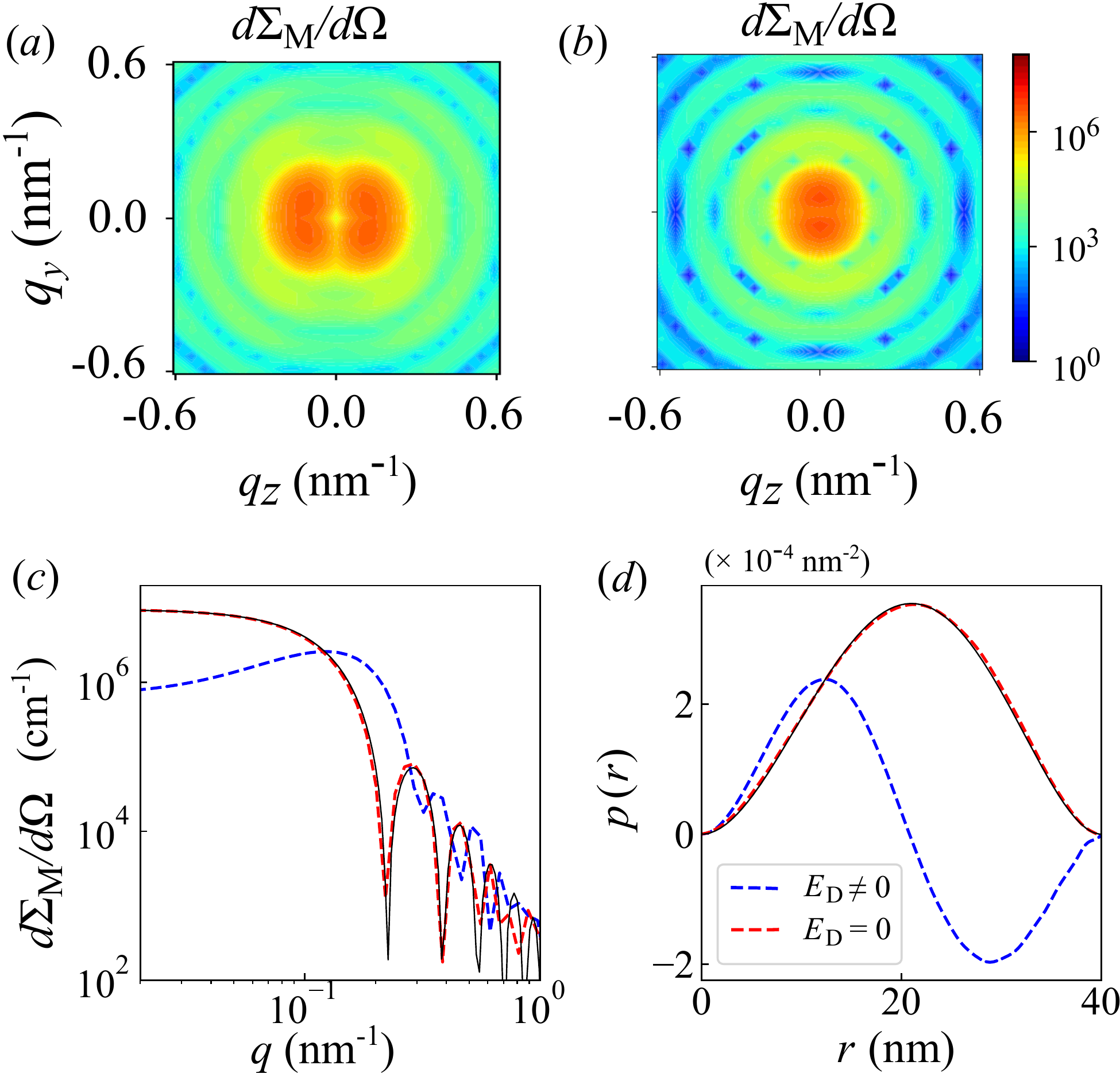}}
\caption{Effect of the dipolar interaction on the SANS observables for $D = 40 \, \mathrm{nm}$, $x_{\mathrm{d}} = 2 \, \%$, and $\mu_0 H_0 = 0 \, \mathrm{T}$. (a)~$d\Sigma_{\mathrm{M}} / d\Omega$ with $E_{\mathrm{D}}$ (logarithmic color scale), (b)~$d\Sigma_{\mathrm{M}} / d\Omega$ for $E_{\mathrm{D}}=0$ (logarithmic color scale), (c)~azimuthally-averaged $d\Sigma_{\mathrm{M}} / d\Omega$ (log-log scale), and (d)~pair-distance distribution function $p(r)$. The thin black lines in (c) and (d) are the analytical results for uniformly magnetized spheres [Eqs.~(\ref{homomagsans2}) and (\ref{pvonreq})].}
\label{fig6}
\end{figure*}

The angular anisotropy of the magnetization Fourier components is caused by the dipolar interaction~\cite{erokhin2012prb}, which is a long-range, nonlocal, and anisotropic magnetic energy term. Figure~\ref{fig6} compares, at remanence, results for $d\Sigma_{\mathrm{M}} / d\Omega$ with and without the dipolar energy $E_{\mathrm{D}}$. It is seen that $d\Sigma_{\mathrm{M}} / d\Omega$ is highly anisotropic (elongated along the horizontal direction) when $E_{\mathrm{D}}$ is included in the computations [Fig.~\ref{fig6}(a)], while it becomes weakly anisotropic (slightly elongated along the vertical direction) when $E_{\mathrm{D}}=0$ [Fig.~\ref{fig6}(b)]. For $E_{\mathrm{D}}=0$, the spin structure remains essentially uniform throughout the magnetization process (compare inset in Fig.~\ref{fig4}(a) and related text) and one observes the analytical sphere form factor results (thin black lines) for the azimuthally-averaged $d \Sigma_{\mathrm{M}} / d \Omega$ [Fig.~\ref{fig6}(c)] and the pair-distance distribution function $p(r)$ [Fig.~\ref{fig6}(d)]. The results in Fig.~\ref{fig6} emphasize the importance of considering complex dipolar-field-induced nonuniform spin textures for the understanding of magnetic SANS patterns. We emphasize, however, that the dipolar energy might be of minor relevance for smaller-sized (nearly uniformly magnetized) nanomagnets and in the presence of the Dzyaloshinskii-Moriya interaction, which may give rise to flux-closure-type magnetization patterns~\cite{hertel2021,koehlerjac2021}.

\begin{figure*}[tb!]
\centering
\resizebox{1.0\columnwidth}{!}{\includegraphics{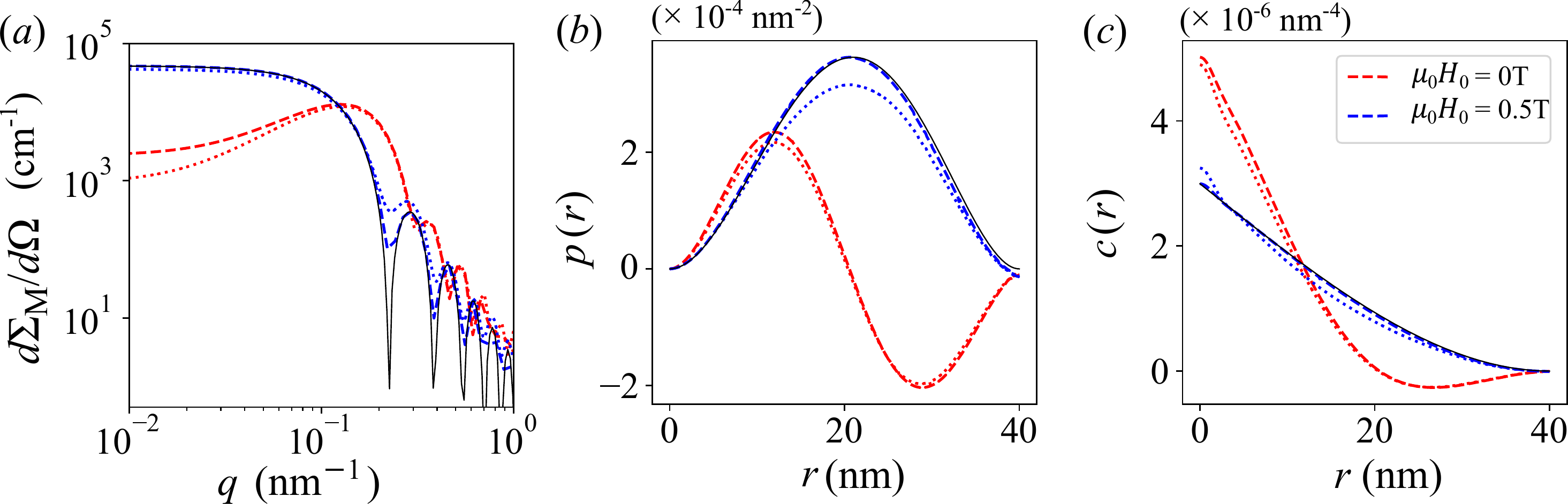}}
\caption{(a)~Azimuthally-averaged magnetic SANS cross section $d\Sigma_{\mathrm{M}} / d\Omega$ (log-log scale), (b)~pair-distance distribution function $p(r)$, and (c)~correlation function $c(r)$ for different applied magnetic fields [$0 \, \mathrm{T}$ and $0.5 \, \mathrm{T}$, see inset in (c)] and defect concentrations. Dashed lines are for $x_{\mathrm{d}} = 0 \, \%$, dotted lines are for $x_{\mathrm{d}} = 15 \, \%$, and the thin black lines correspond to the analytically-known defect-free uniform case [azimuthally-averaged version of Eq.~(\ref{homomagsans2}) for $d\Sigma_{\mathrm{M}} / d\Omega$ and Eqs.~(\ref{pvonreq}) and (\ref{cvonreq}) for, respectively, $p(r)$ and $c(r)$].}
\label{fig7}
\end{figure*}

\begin{figure*}[tb!]
\centering
\resizebox{1.0\columnwidth}{!}{\includegraphics{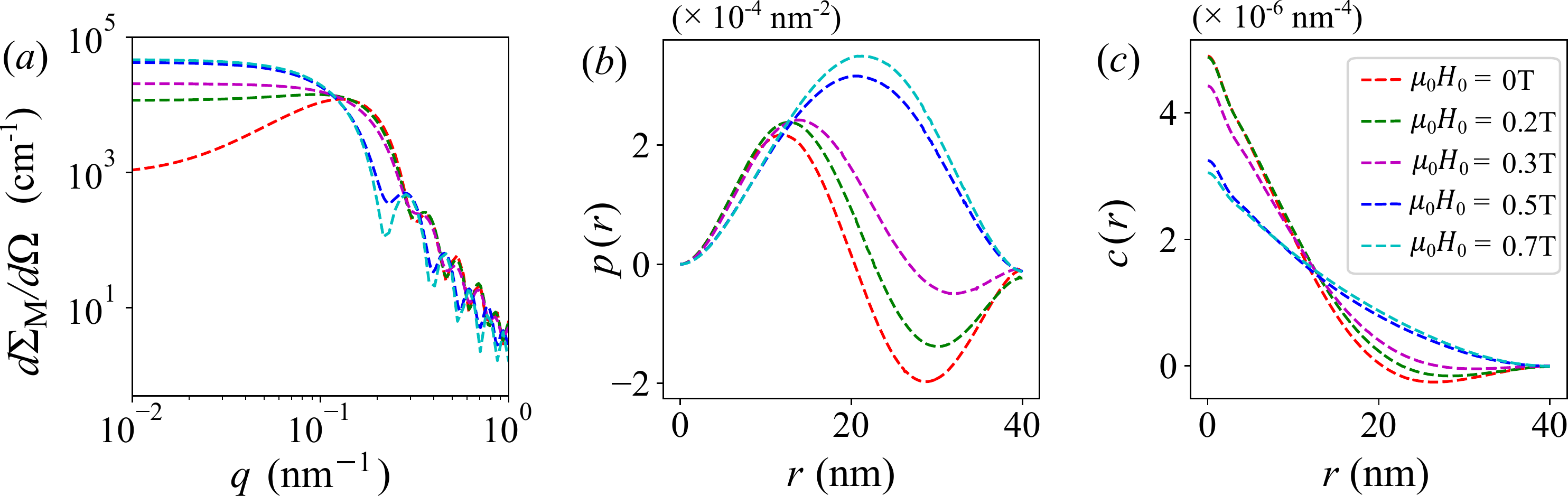}}
\caption{Field dependence [see inset in (c)] of (a)~$d\Sigma_{\mathrm{M}} / d\Omega$ (log-log scale), (b)~$p(r)$, and (c)~$c(r)$ for a defect concentration of $x_{\mathrm{d}} = 15 \, \%$.}
\label{fig8}
\end{figure*}

The results for the ($2\pi$)~azimuthally-averaged $d\Sigma_{\mathrm{M}} / d\Omega$ and for the correlation functions $p(r)$ and $c(r)$ are displayed in Fig.~\ref{fig7} and corroborate the behavior found from the analysis of the two-dimensional SANS cross sections, namely a weak dependence on $x_{\mathrm{d}}$. The signature of the vortex-type real-space spin structure in Fig.~\ref{fig3}(a) is an oscillatory $p(r)$ [see, e.g., Figs.~\ref{fig6}(d) and \ref{fig7}(b)], which appears to be rather stable against spin perturbations that are induced by hole-type defects. Related to that observation is the absence of a Guinier behavior at low momentum transfers $q$ and for small fields [see, e.g., Fig.~\ref{fig7}(a)]. Only with increasing field strength (more uniform spin structure) is a Guinier-type behavior recovered. This becomes also visible in Fig.~\ref{fig8}(a), where the field dependence of $d\Sigma_{\mathrm{M}} / d\Omega$ is shown for $x_{\mathrm{d}} = 15 \, \%$. The emergence of internal spin disorder leads to a shift of the characteristic form-factor oscillations to larger $q$ (smaller structures) and to the smearing of these features, in this way mimicking the effect of a particle-size distribution and/or instrumental resolution [compare also to Figs.~\ref{fig6}(c) and \ref{fig7}(a)]. With increasing field, the $p(r)$ and $c(r)$ in Fig.~\ref{fig8} approach the analytical expressions for uniformly magnetized spheres [Eqs.~(\ref{pvonreq}) and (\ref{cvonreq})]. However, regarding the last statement, one should be cautious and keep in mind that the microstructure of the defect-rich nanoparticles resembles a porous structure (defect cells have a volume of $2 \times 2 \times 2 \, \mathrm{nm^3}$, see Fig.~\ref{fig2}). Such magnetic holes represent a severe defect in the microstructure with a large jump in the saturation magnetization and associated stray-field torques producing spin disorder. Therefore, at large fields when the defect-rich nanoparticles are approaching a uniform magnetization state, it is not surprising that their $c(r)$ at small and intermediate $r$ deviate slightly from the purely uniform case [compare Figs.~\ref{fig7}(c) and \ref{fig8}(c)]. In particular the feature at small $r$ is better resolved in $c(r)$ than in $p(r)$; compare to Fig.~\ref{fig7}(b), where at $0.5 \, \mathrm{T}$ the overall $p(r)$~shape is preserved and only the maximum is reduced at increased $x_{\mathrm{d}}$.

The behavior of the correlation function $c(r)$ in the limit of $r \rightarrow 0$ is generally an interesting question, which provides information on the nature of the scattering contrast (see the discussion by Ciccariello, Goodisman, and Brumberger~\cite{ciccariello88}). When sharp interfaces in a sample separate homogeneous regions with a uniform (constant) scattering length density, then the correlation function exhibits a finite slope at the origin (compare Eqs.~(\ref{pvonreq}) and (\ref{cvonreq}) for uniformly magnetized nanoparticles). This is the content of the well-known Porod law, which predicts a characteristic asymptotic $q^{-4}$~dependency of the SANS cross section~\cite{porod}. By contrast, structures with a nonuniform scattering length density profile, such as the smoothly varying magnetization profiles $\mathbf{M}(\mathbf{r})$ of micromagnetics, are characterized by a $c(r)$ that exhibits a zero slope at $r=0$ and concomitant steeper power-law exponents of the SANS cross section~\cite{michelsbook}. In the present case, we are dealing with a system with a potentially nonuniform magnetization on one side of the interface (inside the particle) and a zero magnetization on the other side of the interface. The behavior of $c(r)$ at small distances also depends on the spin distribution in the vicinity of the surface and, therefore, on the surface anisotropy and the related boundary conditions for the magnetization. This question deserves a separate consideration and is beyond the scope of the present paper.

\subsection{Effect of a particle-size distribution function \label{ressecb}}

In SANS experiments on nanoparticles one always has to deal with a distribution of particle sizes and shapes. The size of a particle has an important effect on its spin structure, e.g., smaller particles generally tend to be uniformly magnetized, whereas larger particles may exhibit inhomogeneous spin structures~\cite{bersweiler2019}. It is therefore also of interest to study the influence of a distribution of particle sizes on the magnetic SANS observables [$d\Sigma_{\mathrm{M}} / d\Omega$, $p(r)$, $c(r)$]. This has been done using a lognormal distribution function, which is defined as~\cite{krill98}:
\begin{eqnarray}
f(D) = \frac{1}{\sqrt{2\pi} D \log\sigma} e^{\displaystyle -\frac{1}{2}\left( \frac{\log D - \log D_0}{\log\sigma} \right)^2} ,
\label{lognorfunc}
\end{eqnarray}
where $D_0$ denotes the median and $\sigma$ the variance of the distribution with $\int_0^{\infty} f(D) dD = 1$. For the above function, the mean particle size $\overline{D}$ [first moment of $f(D)$] is related to the parameters of the distribution as $\overline{D} = D_0 e^\frac{(\log\sigma)^2}{2}$. For given $x_{\mathrm{d}}$ and $B_0$, randomly-averaged magnetic SANS cross sections were computed for particle diameters $D$ ranging between $10 \, \mathrm{nm} \leq D \leq 100 \, \mathrm{nm}$ in binning intervals of $\Delta D = 2 \, \mathrm{nm}$. The magnetic SANS cross section averaged over the distribution, $\langle d\Sigma_{\mathrm{M}} / d\Omega \rangle_f$, is then computed as:
\begin{eqnarray}
\left \langle \frac{d\Sigma_{\mathrm{M}}}{d\Omega} \right \rangle_f = \sum_k w_k \frac{d\Sigma_{\mathrm{M},k}}{d\Omega} ,
\label{lognorfunc}
\end{eqnarray}
where
\begin{eqnarray}
w_k = \int\limits_{D_k - \Delta D / 2}^{D_k + \Delta D / 2} f(D) dD
\label{lognorfuncweigths}
\end{eqnarray}
denotes the weight of the size class $D_k$, which can be computed for given values of $D_0$ and $\sigma$; $d\Sigma_{\mathrm{M},k} / d\Omega$ is the orientationally-averaged SANS cross section [compare Eq.~(\ref{sigmaaverage})] corresponding to $D_k$. Particle diameters outside of the above interval, i.e., smaller than $10 \, \mathrm{nm}$ and larger than $100 \, \mathrm{nm}$ were not considered in our analysis ($w_k=0$ for $D < 10 \, \mathrm{nm}$ and $D > 100 \, \mathrm{nm}$).

\begin{figure}[tb!]
\centering\resizebox{1.0\columnwidth}{!}{\includegraphics{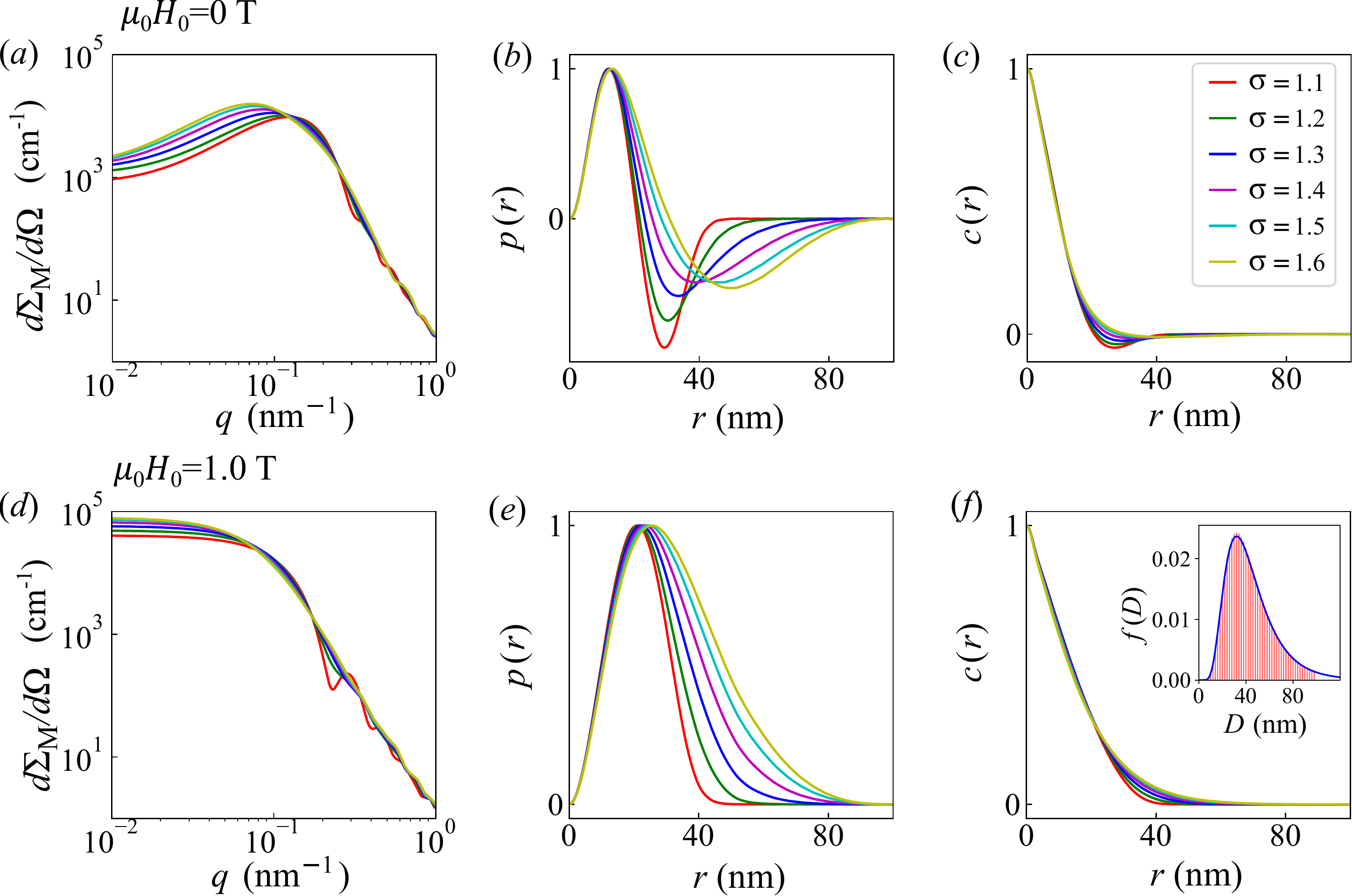}}
\caption{Effect of a lognormal particle-size distribution on the azimuthally-averaged magnetic SANS cross section $d\Sigma_{\mathrm{M}} / d\Omega$ and on the correlation functions $p(r)$ and $c(r)$. Upper row [(a)$-$(c)] is for $\mu_0 H_0 = 0 \, \mathrm{T}$ and the lower row [(d)$-$(f)] is for $\mu_0 H_0 = 1.0 \, \mathrm{T}$. Parameters are:~$D_0 = 40 \, \mathrm{nm}$, $x_{\mathrm{d}} = 0 \, \%$, and $\sigma$ varies between $1.1$$-$$1.6$ [see inset in (c)]. For each $\sigma$, the $p(r)$ and $c(r)$ were normalized to their respective maximum value at $0 \, \mathrm{T}$ and $1.0 \, \mathrm{T}$. The inset in (f) depicts the used size-distribution function $f(D)$ for $D_0 = 40 \, \mathrm{nm}$ and $\sigma = 1.6$. For each size class, 40 random particle orientations were used to compute the averaged magnetic SANS cross section.}
\label{fig9}
\end{figure}

Figure~\ref{fig9} depicts the evolution of the azimuthally-averaged magnetic SANS cross section $d\Sigma_{\mathrm{M}} / d\Omega$ and of both correlation functions $p(r)$ and $c(r)$ with the width $\sigma$ of the lognormal distribution at zero field and at $1.0 \, \mathrm{T}$ (for $D_0 = 40 \, \mathrm{nm}$ and $x_{\mathrm{d}} = 0 \, \%$). Note that the $p(r)$ and $c(r)$ are (for each $D_0$ and $\sigma$) normalized to unity after the cross section has been computed according to Eq.~(\ref{lognorfunc}). The $d\Sigma_{\mathrm{M}} / d\Omega$ [Figure~\ref{fig9}(a)] exhibit the ``usual'' behavior known e.g.\ from the study of instrumental broadening, namely a smearing of the form-factor oscillations with increasing $\sigma$. The remaining oscillations of $d\Sigma_{\mathrm{M}} / d\Omega$ for small $\sigma$ are more pronounced in the high-field regime [Fig.~\ref{fig9}(d)] than at remanence [Fig.~\ref{fig9}(a)]. It is generally seen in Fig.~\ref{fig9}(b) and (e) that the $p(r)$ are more affected by the variation of $\sigma$ than the $c(r)$ [Fig.~\ref{fig9}(c) and (f)], which is related to the $r^2$~factor. For all values of $\sigma$ does the oscillatory $p(r)$~behavior remain at zero field, and one observes a shift of the maximum of $d\Sigma_{\mathrm{M}} / d\Omega$ to lower momentum transfers with increasing $\sigma$ [Fig.~\ref{fig9}(a)]. The global minimum of $p(r)$ at zero field shifts to larger distances with increasing $\sigma$, from $r \cong 29 \, \mathrm{nm}$ for $\sigma = 1.1$ to $r \cong 50 \, \mathrm{nm}$ for $\sigma = 1.6$. For $x_{\mathrm{d}} \neq 0$, the behavior of the $d\Sigma_{\mathrm{M}} / d\Omega$, $p(r)$, and $c(r)$ are qualitatively similar, demonstrating the rather robust character of the oscillatory low-field feature in $p(r)$.

\section{Conclusion and outlook}
\label{summary}

Using micromagnetic computations we have investigated the effect of pore-type microstructural defects in spherical magnetic nanoparticles on their magnetic small-angle neutron scattering cross section $d \Sigma_{\mathrm{M}} / d \Omega$ and pair-distance distribution function $p(r)$. The simulations take into account the isotropic exchange interaction, the magnetocrystalline anisotropy, the dipolar interaction, and an externally applied magnetic field. Clearly, the $d \Sigma_{\mathrm{M}} / d \Omega$ and $p(r)$ of nonuniformly magnetized nanoparticles cannot be described anymore with the superspin model, which assumes a homogeneous spin microstructure. The dipolar interaction is at the origin of many complex magnetization structures and related anisotropic scattering patterns. For small applied fields and a not too small particle size (here, for $D \gtrsim 20 \, \mathrm{nm}$), the dipolar energy results in a vortex-type spin structure and in a concomitant oscillatory feature in the $p(r)$~function. This characteristic signature appears to be rather stable against the here-used pore-type defects. The oscillatory $p(r)$~shape also remains in the presence of a particle-size distribution function. At low fields, deviations from the Guinier law and complicated real-space correlations are encountered. Within the present modeling approach of defects---representing them by computational cells with zero saturation magnetization $M_{\mathrm{s}}$---their effect on the SANS observables seems to be relatively small. The dominating defect in spherical nanoparticles appears to be the outer surface of the particles. A more realistic treatment of defects could be achieved by a more modest change in the material parameters, i.e., by reducing the saturation magnetization to a lower, but nonzero value and/or reducing the exchange interaction between the defect and the other cells~\cite{leliaertjap2014}. Moreover, one could consider the inclusion of the magnetoelastic interaction into the micromagnetic energy functional. Currently, this interaction is not implemented in most micromagnetic codes, although recent research activities go into this direction~\cite{leliaertore2022}. Likewise, phenomenological models for surface anisotropy such as N\'{e}el anisotropy, which give rise to additional boundary conditions on the surface of the nanomagnet, are also not included in numerical simulations of magnetic SANS. The micromagnetic approach to magnetic SANS consist of finding, by means of magnetic-energy minimization, the three-dimensional vector field of the magnetization $\mathbf{M}(\mathbf{r})$. This represents a paradigm shift and is conceptually very much different than the up-to-now used approach of finding a scalar function which describes the structural saturation-magnetization profile $M_{\mathrm{s}}(\mathbf{r})$ of the particle ensemble.

\section*{Acknowledgements}

Evelyn Pratami Sinaga, Michael P.\ Adams, and Andreas Michels acknowledge financial support from the National Research Fund of Luxembourg (PRIDE MASSENA Grant and AFR Grant No.~15639149). Jonathan Leliaert is supported by the Fonds Wetenschappelijk Onderzoek (FWO-Vlaanderen) with senior postdoctoral research fellowship Bo.~12W7622N. The simulations presented in this paper were carried out using the HPC facilities of the University of Luxembourg (\url{https://hpc.uni.lu}).

\bibliographystyle{apsrev4-2}

\begin{thebibliography}{50}
\bibitem [{eve()}]{evelynprbcomment2022}%
  \BibitemOpen
  \href@noop {} {}\bibinfo {note} {Note that the (linearized) analytical
  theories of magnetic SANS are based on Brown's static equations of
  micromagnetics, whereas numerical micromagnetic SANS simulations are ususally
  based on the time-dependent Landau-Lifshitz-Gilbert equation.}\BibitemShut
  {Stop}%
\end{thebibliography}

%

\end{document}